\documentclass[11pt]{article}
\usepackage{amsmath}
\usepackage{amsfonts}
\usepackage{amssymb}
\usepackage{graphicx}
\usepackage{bm}
\usepackage{color}
\usepackage{amscd}

\def\bea{\begin{eqnarray}}
\def\eea{\end{eqnarray}}

\begin{document}
\begin{center}
\LARGE {\bf Lorentz-diffeomorphism quasi-local conserved charges and Virasoro algebra in Chern-Simons-like theories of gravity }
\end{center}

\begin{center}
{M. R. Setare \footnote{E-mail: rezakord@ipm.ir}\hspace{1mm} ,
H. Adami \footnote{E-mail: hamed.adami@yahoo.com}\hspace{1.5mm} \\
{\small {\em  Department of Science, University of Kurdistan, Sanandaj, Iran.}}}\\

\end{center}

\begin{center}
{\bf{Abstract}}\\
The Chern-Simons-like theories of gravity (CSLTG) are formulated at first order formalism. In this formalism, the derivation of the entropy of a black hole on bifurcation surface, as a quasi-local conserved charge is problematic. In this paper we overcome to these problems by considering the concept of total variation and the Lorentz-Lie derivative. We firstly find an expression for the ADT conserved current in the context of the CSLTG  which is based on the concept of the Killing vector fields. Then, we generalize it to be conserved for all diffeomorphism generators. Thus, we can extract an off-shell conserved charge for any vector field which generates a diffeomorphism. The formalism presented here is based on the concept of quasi-local conserved charges which are off-shell. The charges can be calculated on any codimension two space-like surface surrounding a black hole and the results are independent of the chosen surface. By using the off-shell quasi-local conserved charge, we investigate the Virasoro algebra and find a formula to calculate the central extension term. We apply the formalism to the BTZ black hole solution in the context of the Einstein gravity and the Generalized massive gravity, then we find the eigenvalues of their Virasoro generators as well as the corresponding central charges. Eventually, we calculate the entropy of the BTZ black hole by the Cardy formula and we show that the result exactly matches with the one obtained by the concept of the off-shell conserved charges.
\end{center}

\section{Introduction}
As is well known, the concept of conserved charges of gravity theories is related to the concept of the Noether charges corresponding to the Killing vectors which are admitted by solutions of a theory. There are several
 approaches to obtain the mass and angular momentum of black holes solutions of different gravity theories \cite{2'}-\cite{12'}. According to the Arnowitt,
Deser, and Misner method (ADM method) \cite{1'} one can
obtain the conserved charges of an asymptotically flat spacetime
solution of a general theory of relativity, but this is not a
covariant method. The ADM method has been extended to
include asymptotically AdS spacetime solution of Einstein gravity \cite{2'}. Deser and Tekin have
extended this approach. By this extension one can calculate the energy of asymptotically
dS or AdS solutions in higher curvature gravity
models and also in a topologically massive gravity model (TMG) \cite{3'}. This method is a covariant formalism; it is known as
the Abbott–-Deser-–Tekin (ADT) formalism. Another method is the Brown-York approach \cite{121'}
 which is based on a quasi-local concept, but this approach also
is not covariant. The authors of \cite{15'} have computed the ADT charges for a solution of
TMG linearized about an arbitrary background and have applied the result
to evaluate the mass and angular momentum of the non-asymptotically flat,
non-asymptotically AdS black hole solution (ACL black hole) of TMG.\\
A general definition of conserved charges in general relativity
and other theories of gravity has been proposed in \cite{50}. In the metric formalism of gravity for the covariant theories defined by a Lagrangian $n$-form $L$, Wald has shown that the entropy of black holes is the Noether charge associated with the horizon-generating Killing vector field evaluated at the bifurcation surface. Tachikawa extended the Wald approach to include non-covariant theories \cite{100'}. Hence, regarding this extension one can calculate the black hole entropy as a Noether charge in the context of non-covariant theories as well.
 But it is clear now that the derivation of the classical Wald formula for entropy is problematic in the first order formalism using the spin connection. In the first order formalism, the expression of conserved charges are proportional to the Killing vector field $\xi$.
   It is clear that $\xi$ must be zero on the bifurcation surface when we calculate the entropy of black hole, because $\xi$ is the horizon-generating Killing vector field which is zero on the bifurcation surface. It seems disappointing at the first glance because it appears that the entropy will be zero, but it is not true.
  In the other hand there is a class of of gravitational theories in $(2+1)$-dimension (e.g. Topological massive gravity (TMG) \cite{111'}, New massive gravity (NMG)\cite{211'}, Generalized Massive Gravity (GMG) \cite{17}, Minimal massive gravity (MMG) \cite{311'}, Zewi-dreibein gravity (ZDG)\cite{411'}, Generalized minimal massive gravity (GMMG) \cite{511'}, e.t.c) which well-known as the Chern-Simons-like theories of gravity(CSLTG) \cite{1}, can be written in the first order formalism. The authors of \cite{3} have shown that in approaching to the bifurcation surface, the spin-connection diverges in a way that the spin-connection interior product in $\xi$  remains finite ensuring that there is no problem. Recently we have extended this formalism in the on-shell case, to the Lorentz-diffeomorphism non-covariant theories \cite{2}. Here we would like extend this formalism to the off-shell case in the framework of CSLTG. We derive the conserved charge formula by a new method and explicitly compute conserved charges in some models.\\
The authors of \cite{5} have obtained
the quasi-local conserved charges for black holes in any diffeomorphically
invariant theory of gravity. By considering an appropriate variation of the
metric, they have established a one-to-one correspondence between the ADT
approach and the linear Noether expressions.\footnote{In quasi-local approach \cite{5,6}, the Killing vector field $\xi$ is defined  not only in the asymptotical part of space, but also  in any other points of the space-time. For example see equation (33) below, where $\Sigma$ is an arbitrary space-like codimension-two surface. As can be seen in Eq.(22), $\mathcal{J}_{ADT}$ is conserved current when $\xi$ is a Killing vector field in any point of space-time. In this case $\delta _{\xi} a^{r} = \delta _{\xi} E_{r}=0$, so $d\mathcal{J}_{ADT}=0$. } They have extended this
work to a theory of gravity containing a gravitational Chern-Simons term
in \cite{6}, and have computed the off-shell potential and quasi-local conserved
charges of some black holes in TMG. We have obtained  the quasi-local conserved charges of Lorentz-diffeomorphism covariant theories of gravity in the first order formalism, in paper \cite{29}. But there are theories which are not Lorentz-difeomorphism covariant.  In previous paper \cite{2} by introducing the total variation of a quantity due to the infinitesimal Lorentz-diffeomorphism transformation, we have obtained the conserved charges in the Lorentz-diffeomorphism non-covariant
theories. The proposed formalism in \cite{2} is for on-shell case. As we have mentioned in the above, here we are going to generalize the proposed formalism to the off-shell case. We try to find an expression for the ADT conserved current which is off-shell conserved for a given Killing vector field. We generalize off-shell conserved current $\mathcal{J}_{ADT}$ to make sure that it is conserved for any diffeomorphism generator $\xi$. For this purpose, we follow the method presented in \cite{7} in which the authors took advantage of the metric formalism. $\mathcal{J}_{ADT}$ is  off-shell conserved if $\xi$ is a Killing vector field. We will show that one can introduce the generalized ADT current $\tilde{\mathcal{J}} _{ADT}$ which is off-sell conserved for an arbitrary diffeomorphism generator $\xi$. Then, we can find the generalized ADT
conserved charge by virtue of the Poincare lemma, such that $ \tilde{\mathcal{J}} _{ADT} = d \tilde{\mathcal{Q}} _{ADT} $. We  will fix the ambiguity in definition of the generalized ADT conserved current by considering the phase space
analysis. Then we try to obtain the central extension term, $C (\xi _{m} ^{\pm} , \xi _{n} ^{\pm})$, for the algebra of the conserved charges in the context of the CSLTG. We apply our formalism to the Einstein gravity in the presence of negative cosmological constant, and also to the GMG. We will find the central charges of dual CFT of the BTZ black hole solutions of mentioned gravity theories. Then by obtaining the eigenvalues of the Virasoro generators, $l^{\pm} _{n}$, and using the Cardy formula we will obtain the Bekenstein-Hawking entropy of the BTZ black hole, as well as the energy and the angular-momentum of the BTZ solution of GMG. By using the Killing vectors corresponding to the mentioned quantities we will obtain energy, angular-momentum  and entropy of the BTZ black hole solution of GMG again which are exactly coincide with previous results.
\section{Generalized conserved current}
In this section, we are going to find an off-shell conserved current and corresponding conserved charge of the Chern-Simons-like theories of gravity. We generalize this conserved current so that corresponds to an arbitrary diffeomorphism rather than a diffeomorphism which is generated by a Killing vector field. \\
The Lagrangian 3-form of the Chern-Simons-like theories of gravity ( CSLTG ) is given by \cite{1}
\begin{equation}\label{1}
  L=\frac{1}{2} g_{rs} a^{r} \cdot da^{s}+\frac{1}{6} f_{rst} a^{r} \cdot a^{s} \times a^{t}.
\end{equation}
In the above Lagrangian $ a^{ra}=a^{ra}_{\hspace{3 mm} \mu} dx^{\mu} $ are Lorentz vector valued one-forms where, $r=1,...,N$ and $a$ indices are refer to flavour and Lorentz indices, respectively. We should mention that, here, the wedge products of Lorentz-vector valued one-form fields is implicit. Also, $g_{rs}$ is a symmetric constant metric on the flavour space and $f_{rst}$ is a totally symmetric "flavour tensor" which interpret as the coupling constants. We use a 3D-vector algebra notation for Lorentz vectors in which contractions with $\eta _{ab}$ and $\varepsilon ^{abc}$ are denoted by dots and crosses, respectively \footnote{Here we consider the notation used in \cite{1}.}. We know that $a^{ra}$ is a collection of the dreibein $e^{a}$, the dualized spin-connection $\omega ^{a}$, the auxiliary field $ h^{a}_{\hspace{1.5 mm} \mu} = e^{a}_{\hspace{1.5 mm} \nu} h^{\nu}_{\hspace{1.5 mm} \mu} $ and so on. It is worth mention that for all interesting CSLTG we have $f_{\omega rs} = g_{rs}$ \cite{4}. \\
The total variation of $a^{ra}$ due to a diffeomorphism generator $\xi$ is \cite{2}
\begin{equation}\label{2}
  \delta _{\xi} a^{ra} = \mathfrak{L}_{\xi} a^{ra} -\delta ^{r} _{\omega} d \chi _{\xi} ^{a} ,
\end{equation}
which is caused by a combination of variations due to the diffeomorphism and the infinitesimal Lorentz gauge transformation. In the above expression $\mathfrak{L}_{\xi}$ denotes the Lie-Lorentz derivative (L-L derivative) which is defined by \cite{3}
\begin{equation}\label{3}
  \mathfrak{L}_{\xi} A^{a \cdots } _{ \hspace{4.5 mm} b \cdots } = \pounds_{\xi} A^{a \cdots } _{ \hspace{4.5 mm} b \cdots } + \lambda ^{a}_{ \xi \hspace{1 mm} c} A^{c \cdots } _{ \hspace{4.5 mm} b \cdots } + \cdots - \lambda ^{c}_{\xi \hspace{1 mm} b} A^{a \cdots } _{ \hspace{4.5 mm} c \cdots } - \cdots ,
\end{equation}
where $\pounds_{\xi}$ denotes the ordinary Lie derivative along $\xi$ and $\lambda ^{ab}_{\xi}$ generates the Lorentz gauge transformations $SO(n-1,1)$. In general, $\lambda ^{ab}_{\xi}$ is a general function of space-time coordinates and of the diffeomorphism generator $\xi$. Also, $\chi _{\xi} ^{a}$ is defined as $\chi _{\xi} ^{a}= \frac{1}{2} \varepsilon ^{a} _{\hspace{1.5 mm} bc} \lambda _{\xi}^{bc} $ and $ \delta ^{r} _{s} $  denotes the ordinary Kronecker delta. It should be noted that $\chi _{\xi}$ is linear in term of $\xi$. Presence of the extra term , $-\delta ^{r} _{\omega} d \chi _{\xi} ^{a}$, in \eqref{2} may be caused the Lagrangian \eqref{1} do not invariant under a general Lorentz-diffeomorphism transformation. \\
The variation of the Lagrangian \eqref{1} is given by
\begin{equation}\label{4}
  \delta L = \delta a^{r} \cdot E_{r} + d \Theta (a, \delta a) ,
\end{equation}
where
\begin{equation}\label{5}
   E_{r}^{\hspace{1.5 mm} a} = g_{rs} d a^{sa} + \frac{1}{2} f_{rst} (a^{s} \times a^{t})^{a} ,
\end{equation}
so that $E_{r}^{\hspace{1.5 mm} a}=0$ are the equations of motion, and
\begin{equation}\label{6}
   \Theta (a, \delta a) = \frac{1}{2} g_{rs} \delta a^{r} \cdot a^{s},
\end{equation}
is the surface term.
By virtue of \eqref{2}, the total variation of the Lagrangian due to diffeomorphism generator $ \xi $ can be written as
\begin{equation}\label{7}
  \delta _{\xi} L = \mathfrak{L}_{\xi} L +d \psi _{\xi} ,
\end{equation}
where $\psi _{\xi}$ is given by
\begin{equation}\label{8}
   \psi _{\xi} = \frac{1}{2} g_{\omega r} d \chi _{\xi} \cdot a^{r} .
\end{equation}
Now, by considering that the variation in \eqref{4} be the total variation generated by $\xi$ and using \eqref{7}, we will have
\begin{equation}\label{9}
\begin{split}
   d J = & (i_{\xi} \omega - \chi _{\xi}) \cdot ( D E_{\omega} + a ^{r^{\prime}} \times E_{r^{\prime}} ) \\
         & + i_{\xi} a ^{r^{\prime}} \cdot D E_{r^{\prime}} - i_{\xi} D a ^{r^{\prime}} \cdot E_{r^{\prime}} -i_{\xi} R \cdot E_{\omega} ,
\end{split}
\end{equation}
where, we define $J$ as follows:
\begin{equation}\label{10}
  J (\xi ) = \Theta (a, \delta _{\xi} a ) - i_{ \xi } L - \psi _{\xi} + i_{\xi} a^{r} \cdot E _{r} - \chi _{\xi} \cdot E _{\omega}.
\end{equation}
In the above equations $i _{\xi}$, $D$ and $R=d \omega + \frac{1}{2} \omega \times \omega $ are interior product, exterior covariant derivative and the curvature 2-form, respectively. Also, the prime on $r$ indicates that the sum run over all the flavour indices except $\omega$.\\
We expect that the last line in \eqref{9} can be rewritten as
\begin{equation}\label{11}
 i_{\xi} a ^{r^{\prime}} \cdot D E_{r^{\prime}} - i_{\xi} D a ^{r^{\prime}} \cdot E_{r^{\prime}} -i_{\xi} R \cdot E_{\omega} = i_{\xi} a ^{r^{\prime}} \cdot X _{r^{\prime}}(a) .
\end{equation}
The Bianchi identities can be expressed as following identities
\begin{equation}\label{12}
  D E_{\omega} + a ^{r^{\prime}} \times E_{r^{\prime}}=0, \hspace{1.5 cm} X _{r^{\prime}}^{\hspace{1.5 mm} a}(a^{s}) =0 .
\end{equation}
To clarifying this, one can consider the Einstein gravity with the cosmological term in which we have
\begin{equation}\label{13}
  a^{r}=\{ e, \omega \} , \hspace{1 cm} g_{e \omega} = -1 , \hspace{1 cm} f _{e \omega \omega}=-1 , \hspace{1 cm} f _{eee}= \Lambda .
\end{equation}
In this theory, the Bianchi identities introduced in \eqref{12} reduce to the ordinary Bianchi identities respectively
\begin{equation}\label{14}
  DT=R \times e , \hspace{1 cm} DR=0 .
\end{equation}
Due to the Bianchi identities \eqref{12}, $ J(\xi ) $ is an closed form and, by the Poincare lemma, it can be written as $J (\xi ) = d K (\xi )$. \\
Now, we take an arbitrary variation over $ J$ in \eqref{10} then we obtain
\begin{equation}\label{15}
\begin{split}
   d \left( \delta K ( \xi ) - i_{\xi} \Theta (a , \delta a )\right) = & \delta \Theta (a , \delta _{\xi} a ) - \mathfrak{L}_{\xi} \Theta (a , \delta a ) - \delta \psi _{\xi}  \\
     & + \delta a^{r} \cdot i_{\xi} E_{r} + i_{\xi} a^{r} \cdot \delta E_{r} - \chi _{\xi} \cdot \delta E_{\omega}.
\end{split}
\end{equation}
On the one hand, the total variation of the surface term $\Theta (a , \delta a )$ is given by
\begin{equation}\label{16}
\delta _{\xi} \Theta (a , \delta a ) = \mathfrak{L}_{\xi} \Theta (a , \delta a ) + \Pi _{\xi},
\end{equation}
where
\begin{equation}\label{17}
\Pi _{\xi}= \frac{1}{2} g_{\omega r} d \chi _{\xi} \cdot \delta a^{r} .
\end{equation}
On the other hand, the symplectic current is defined as
\begin{equation}\label{18}
     \Omega (a , \delta _{1} a , \delta _{2} a )  = \delta _{1} \Theta (a , \delta _{2} a) - \delta  _{2} \Theta (a, \delta _{1} a).
\end{equation}
Using the equations (\ref{8}), (\ref{16}), (\ref{17}), and (\ref{18}), the equation \eqref{15} can be rewritten as
\begin{equation}\label{19}
   d \left( \delta K ( \xi ) - i_{\xi} \Theta (a , \delta a ) \right) =  \Omega (a, \delta a , \delta _{\xi} a ) + \delta a^{r} \cdot i_{\xi} E_{r} + i_{\xi} a^{r} \cdot \delta E_{r} - \chi _{\xi} \cdot \delta E_{\omega}.
\end{equation}
 Since the symplectic current is linear in $\delta _{\xi} a$, when $\xi$ is a Killing vector filed the symplectic current vanishes. So, in the manner of the papers \cite{5,6}, we can consider the ADT conserved current and the ADT conserved charge as \footnote{Regarded to the Eq.\eqref{20}, in fact we have purposed this equation follow from Eq.(5) in \cite{5} (see also \cite{6,15',12'}). In the another term, Eq.\eqref{20} in our paper is first- order analog of Eq.(5) in \cite{5}, which has written in metric formalism. Eq.\eqref{21} come from Eqs.(19), (20) easily. Also from Eq.\eqref{22} it is clear that when $\xi$ is Killing vector field in any given point of space time, the current $ \mathcal{J}_{ADT}$ is conserved current.}
\begin{equation}\label{20}
    \mathcal{J}_{ADT} = \delta a^{r} \cdot i_{\xi} E_{r} + i_{\xi} a^{r} \cdot \delta E_{r} - \chi _{\xi} \cdot \delta E_{\omega},
\end{equation}
and
\begin{equation}\label{21}
   \mathcal{Q} _{ADT} (\delta a, \xi) = \delta K ( \xi ) - i_{\xi} \Theta (a , \delta a ),
\end{equation}
respectively. The last term in \eqref{20} help us to define a covariant conserved current and consequently a covariant conserved charge. \\
As we know, $\mathcal{J}_{ADT}$ and $\mathcal{Q} _{ADT}$ are off-shell conserved current and off-shell conserved charge , respectively, only when $ \xi $ is a Killing vector field. Now, we try to generalize $\mathcal{J}_{ADT}$ such that it is not only off-shell conserved but it becomes conserved for any diffeomorphism generator $\xi $ rather than a Killing vector filed diffeomorphism generator. For this purpose, we follow the method presented in the paper \cite{7} which has written in the metric formalism. To this end, we take an exterior derivative form $\mathcal{J}_{ADT}$ \eqref{20} and we find following relation
\begin{equation}\label{22}
   d \mathcal{J}_{ADT} = \delta _{\xi} a^{r} \cdot \delta E_{r} - \delta a^{r} \cdot \delta _{\xi} E_{r},
\end{equation}
which is first order analogue of Eq.(2.6) in \cite{7}.
From the above equation, it is obvious that $\mathcal{J}_{ADT}$ is conserved off-shell if $\xi$ be a Killing vector field. The right hand side of the above equation can be rewritten as an exact form
\begin{equation}\label{23}
    \delta _{\xi} a^{r} \cdot \delta E_{r} - \delta a^{r} \cdot \delta _{\xi} E_{r} = - d \left( g_{rs} \delta _{\xi} a^{r} \cdot \delta a^{s} \right) = - d \mathcal{J}_{\Delta} .
\end{equation}
Thus, we can introduce the generalized ADT current as
\begin{equation}\label{24}
  \tilde{\mathcal{J}} _{ADT} = \mathcal{J}_{ADT} + \mathcal{J}_{\Delta} .
\end{equation}
so that $d \tilde{\mathcal{J}} _{ADT} =0$ and hence, it is off-sell conserved for an arbitrary diffeomorphism generator $\xi $. In this way, we can find the generalized ADT conserved charge by virtue of the Poincare lemma, $\tilde{\mathcal{J}} _{ADT} = d \tilde{\mathcal{Q}} _{ADT} (\delta a, \xi) $. It is obvious that \eqref{24} reduces to the ordinary ADT conserved current when $\xi$ is a Killing vector field.

\section{Comparison with the covariant phase space analysis}
We know that the variation of the Lagrangian 3-form \eqref{1} is given by \eqref{4}. So, by taking another variation from \eqref{4}, and demanding $\delta _{1} \delta _{2} L = \delta _{2} \delta _{1} L $ and $\delta _{1} \delta _{2} a^{r} = \delta _{2} \delta _{1} a^{r} $, we will have
\begin{equation}\label{25}
  d \Omega (a , \delta _{1} a , \delta _{2} a ) = \delta _{1} a^{r} \cdot \delta _{2} E_{r} - \delta _{2} a^{r} \cdot \delta _{1} E_{r} ,
\end{equation}
where $\Omega (a , \delta _{1} a , \delta _{2} a )$ is the symplectic current which is defined by \eqref{18}. It is clear that the symplectic current is conserved when $a^{r}$ and $\delta a^{r}$ satisfy the equations of motion and the linearized equations of motion respectively. If we take $\delta _{1}= \delta $ and $\delta _{2} = \delta _{\xi}$ then $d \Omega$  will given by
\begin{equation}\label{26}
  d \Omega (a , \delta a , \delta _{\xi} a ) = \delta a^{r} \cdot \delta _{\xi} E_{r} - \delta _{\xi} a^{r} \cdot \delta E_{r} = d \mathcal{J}_{\Delta} ,
\end{equation}
in the last equality we have used \eqref{23}, so
\begin{equation}\label{27}
  \mathcal{J}_{\Delta} = \Omega (a , \delta a , \delta _{\xi} a ) + d Z (a , \delta a , \delta _{\xi} a ) .
\end{equation}
Using Eq.(\ref{6}) and definition of symplectic current \eqref{18}, we obtain
\begin{equation}\label{28}
  \Omega (a , \delta a , \delta _{\xi} a ) = g_{rs} \delta _{\xi} a^{r} \cdot \delta a^{s}.
\end{equation}
By substituting \eqref{28} into \eqref{27} and comparing the obtained result with the last equality in \eqref{23}, we find that $Z (a , \delta a , \delta _{\xi} a )$ can be chosen to be zero. In this way, the generalized ADT current will have the following form
\begin{equation}\label{29}
  \tilde{\mathcal{J}} _{ADT} = \mathcal{J}_{ADT} + \Omega (a , \delta a , \delta _{\xi} a ) .
\end{equation}
From this equation, we easily see that, if $\xi$ be a Killing vector field then the generalized ADT current is reduced to the ordinary one and, if the equations of motion and the linearized equations of motion both satisfied then the ADT current is reduced to the symplectic current, as we expected.

\section{Conserved charges}
In the previous sections, we generalized the ADT conserved current of the CSLTG so that it be conserved for a general diffeomorphism generator $\xi$. In this section, using the concept of the generalized ADT conserved charge, we extract a general formula for the conserved charges. As we know, the generalized ADT current is conserved off-shell and for any diffeomorphism generator $\xi$, $d \tilde{\mathcal{J}} _{ADT} =0 $. Then, by virtue of the Poincare lemma, we can introduce the generalized ADT conserved charge $\tilde{\mathcal{Q}} _{ADT}$ such that, $ \tilde{\mathcal{J}} _{ADT} = d \tilde{\mathcal{Q}} _{ADT} (\delta a , \xi) $. By comparing \eqref{19}, \eqref{20} and \eqref{29}, we deduce that
\begin{equation}\label{30}
   \tilde{\mathcal{Q}} _{ADT} (\delta a , \xi) = \delta K ( \xi ) - i_{\xi} \Theta (a , \delta a ).
\end{equation}
On the other hand, one can read off $K(\xi)$ from \eqref{10} as
\begin{equation}\label{31}
    K ( \xi ) = \frac{1}{2} g_{rs} i_{\xi} a^{r} \cdot a^{s} - g _{\omega s} \chi _{\xi} \cdot a^{s} ,
\end{equation}
so, by substituting \eqref{31} into \eqref{30}, we find that the generalized ADT conserved charge is given by
\begin{equation}\label{32}
   \tilde{\mathcal{Q}} _{ADT} (\delta a , \xi) = \left( g_{rs} i_{\xi} a^{r} - g _{\omega s} \chi _{\xi} \right) \cdot \delta a^{s} .
\end{equation}
Now, we can define the conserved charge perturbation for the diffeomorphism generator $\xi$ as follows:
\begin{equation}\label{33}
   \delta Q ( \xi )  \equiv c \int_{\Sigma} \tilde{\mathcal{Q}} _{ADT} (\delta a , \xi) = c \int_{\Sigma} \left( g_{rs} i_{\xi} a^{r} - g _{\omega s} \chi _{\xi} \right) \cdot \delta a^{s},
\end{equation}
where $\Sigma$ is an arbitrary space-like codimension two surface and $c$ is just a normalization factor. For obtaining the conserved charges of a black hole solution, we can take an integration from \eqref{33} over the one-parameter path on the solution space \cite{5} so we have
\begin{equation}\label{34}
    Q ( \xi )  = c \int_{0}^{1} ds \int_{\Sigma} \tilde{\mathcal{Q}} _{ADT} (s | \xi) .
\end{equation}
It should be noted that \eqref{33} is an off-shell conserved charge perturbation for an arbitrary diffeomorphism generator $\xi$.
Presence of $\chi _{\xi}$ in \eqref{33} is due to the Lorentz gauge transformation. In absence of this term the conserved charge is proportional to $\xi$. Now to obtain entropy as a conserved charge, we should consider $\Sigma$ as a bifurcation surface, then we should put $\xi=0$, which leads to a zero value for entropy. But as it is clear from Eq.(46) in below,  $\chi _{\xi}$ is proportional to  $ \nabla ^{\mu} \xi ^{\nu}$, so the presence of this term give us a correct value for the entropy of black hole. Therefore, using the definition of total variation we could overcome on the mentioned problem.
\section{Virasoro algebra and the central term}
Using the result of the previous section, we can obtain the central charges of the CSLTG. So in this section we find the central extension term for these theories form which one can read off the central charges. We know that, two copies of the classical centerless Virasoro algebra, which is known as the Witt algebra, is given by
\begin{equation}\label{35}
    [ \xi _{m} ^{\pm} , \xi _{n} ^{\pm} ] = i (n-m) \xi _{m+n} ^{\pm} , \hspace{1 cm} [ \xi _{m} ^{+} , \xi _{n} ^{-} ] = 0 ,
\end{equation}
where $\xi _{m} ^{\pm}$ ( $ m \in \mathbb{Z}$ ) are the vector fields and the square brackets denote the Lie bracket. \\
The central extension term $C (\xi _{m} ^{\pm} , \xi _{n} ^{\pm})$ is given by the following equation \cite{8}
\begin{equation}\label{36}
   [ Q ( \xi _{m} ^{\pm} ) , Q ( \xi _{n} ^{\pm} ) ]  = Q ( [ \xi _{m} ^{\pm} , \xi _{n} ^{\pm} ] ) +  C (\xi _{m} ^{\pm} , \xi _{n} ^{\pm}) .
\end{equation}
Since the conserved charge \eqref{33} is linear in $ \xi $, then
\begin{equation}\label{37}
 Q ( [ \xi _{m} ^{\pm} , \xi _{n} ^{\pm} ] ) = i (n-m) Q( \xi _{m+n} ^{\pm} ) ,
\end{equation}
on the other hand, we know that
\begin{equation}\label{38}
   [ Q ( \xi _{m} ^{\pm} ) , Q ( \xi _{n} ^{\pm} ) ]  = \delta _{\xi _{n} ^{\pm}} Q( \xi _{m} ^{\pm} ),
\end{equation}
thus, the central extension term will be obtained from the following equation
\begin{equation}\label{39}
  C (\xi _{m} ^{\pm} , \xi _{n} ^{\pm})   = \delta _{\xi _{n} ^{\pm}} Q( \xi _{m} ^{\pm} ) - i (n-m) Q( \xi _{m+n} ^{\pm} ) .
\end{equation}
 Since for a black hole solution, the integration surface $\Sigma$ can be taken as an arbitrary ($t,r$)-constant surface then, the obtained result will be same for any choice of the integration surface, at the spatial infinity \cite{8,9} or at the near horizon region \cite{10,11}. So, we can take $\Sigma$ at spatial infinity without losing the generality and we are confident that the obtained result holds on the near horizon. \\
Now, at the spatial infinity, we can rewrite the conserved charge as follows:
\begin{equation}\label{40}
    Q ( \xi )  = -\frac{1}{8 \pi G} \int_{\infty} \left( g_{rs} i_{\xi} a^{r} - g _{\omega s} \chi _{\xi} \right) \cdot \delta a^{s},
\end{equation}
where quantities are calculated on the background and $\delta a^{s}$ denotes deviation of the 1-form valued fields form background one. In the above experssion (and from now on) we take the normalization factor as $c=-\frac{1}{8 \pi G}$. In the left hand side of \eqref{40} we are dropped $\delta$ in front of $Q$ because, what we calculate in the right hand side is exactly the conserved charge ( one can deduce this from \eqref{34} and the concept of the spatial infinity). Now, from \eqref{40} and \eqref{33}, we have
\begin{equation}\label{41}
    Q ( \xi _{m} ^{\pm} )  = -\frac{1}{8 \pi G} \int_{\infty} \left( g_{rs} i_{\xi _{m} ^{\pm}} a^{r} - g _{\omega s} \chi _{\xi _{m} ^{\pm}} \right) \cdot \delta a^{s},
\end{equation}
and
\begin{equation}\label{42}
   \delta _{\xi _{m} ^{\pm}} Q ( \xi _{n} ^{\pm} )  = -\frac{1}{8 \pi G} \int_{\infty} \left( g_{rs} i_{\xi _{n} ^{\pm}} a^{r} - g _{\omega s} \chi _{\xi _{n} ^{\pm}} \right) \cdot \delta _{\xi _{m} ^{\pm}} a^{s}.
\end{equation}
Then, by substituting these results in \eqref{39}, we will find an expression for the central extension term and consequently we can read off the central charges of the considered theory.
\section{Examples}
In this section, we apply the obtained results in the previous section on the Einstein gravity with the negative cosmological constant and on the Generalized Massive Gravity for the BTZ black hole solution. We will see that this method is easier and more clear than others method which are proposed until now. At first, we derive some useful equations. \\
The AdS$_{3}$ metric is given by
\begin{equation}\label{43}
   ds^{2}= -\frac{r^{2}}{l^{2}} dt^{2} + \frac{l^{2}}{r^{2}} dr^{2} + r^{2} d \phi ^{2},
\end{equation}
so, we can write down the dreibein as follows:
\begin{equation}\label{44}
   e^{\hat{t}} = \frac{r}{l} dt, \hspace{1 cm} e^{\hat{r}} = \frac{l}{r} dr, \hspace{1 cm} e^{\hat{\phi}} = r d \phi .
\end{equation}
We will take these as the background dreibein. As we know, the following vector fields satisfy the Witt algebra \cite{12}
\begin{equation}\label{45}
   \xi _{n} ^{\pm} = \frac{1}{2} e^{inx^{\pm}} \left[  l \left( 1-\frac{l^{2} n^{2}}{2r^{2}} \right) \partial _{t} -inr \partial _{r} \pm \left( 1+\frac{l^{2} n^{2}}{2r^{2}} \right) \partial _{\phi } \right] ,
\end{equation}
where $x^{\pm}= t/l \pm \phi$. \\
If we demand that $\delta _{\xi} e^{a}= 0$ when $\xi$ is Killing, we find the following expression for $\chi_{\xi}$ \cite{2,3}
\begin{equation}\label{46}
  \chi _{\xi} ^{a} = i_{\xi} \omega ^{a} -\frac{1}{2} \varepsilon ^{a}_{\hspace{1.5 mm} bc} e^{\nu b} (i_{\xi} T^{c})_{\nu} + \frac{1}{2} \varepsilon ^{a}_{\hspace{1.5 mm} bc} e^{b}_{\hspace{1.5 mm} \mu} e^{c}_{\hspace{1.5 mm} \nu} \nabla ^{\mu} \xi ^{\nu} .
\end{equation}
Since the following examples are torsion-free so we will have
\begin{equation}\label{47}
  i_{\xi _{n} ^{\pm}} \omega ^{a} - \chi _{\xi _{n} ^{\pm}} ^{a} = \pm \frac{1}{l} (\xi _{n} ^{\pm})^{a} .
\end{equation}
On the other hand, the total variation of dreibein and the spin-connection along the diffeomorphism generator $\xi$ are \cite{3}
\begin{equation}\label{48}
  \delta _{\xi} e^{a} _{\hspace{1.5 mm} \mu} = \frac{1}{2} e^{a \nu} \pounds _{\xi} g_{\mu \nu}  , \hspace{1 cm} \delta _{\xi} \omega ^{a} = i_{\xi} R^{a} + D ( i_{\xi} \omega ^{a} -\chi _{\xi} ^{a} ) ,
\end{equation}
For the AdS$_{3}$ spacetime we have $R=-\frac{1}{2l^{2}} e \times e$ so we find that
\begin{equation}\label{49}
  \begin{split}
       & \delta _{\xi _{n} ^{\pm}} \omega ^{\hat{t}} _{\hspace{1.5 mm} \phi} \pm \frac{1}{l} \delta _{\xi _{n} ^{\pm}} e^{\hat{t}} _{\hspace{1.5 mm} \phi} = - \frac{i l n^{3}}{2r} e^{inx^{\pm}} , \\
       & \delta _{\xi _{n} ^{\pm}} \omega ^{\hat{r}} _{\hspace{1.5 mm} \phi} \pm \frac{1}{l} \delta _{\xi _{n} ^{\pm}} e^{\hat{r}} _{\hspace{1.5 mm} \phi} = 0 , \\
       & \delta _{\xi _{n} ^{\pm}} \omega ^{\hat{\phi}} _{\hspace{1.5 mm} \phi} \pm \frac{1}{l} \delta _{\xi _{n} ^{\pm}} e^{\hat{\phi}} _{\hspace{1.5 mm} \phi} = \pm \frac{i l n^{3}}{2r} e^{inx^{\pm}}.
  \end{split}
\end{equation}
Now, as mentioned above, we take AdS$_{3}$ as the background spacetime and then we consider the following examples.

\subsection{Einstein gravity with the negative cosmological constant}
The Lagrangian 3-form of the Einstein gravity with the negative cosmological constant is given by
\begin{equation}\label{50}
  L_{E} = - e \cdot R - \frac{1}{6 l^{2}} e \cdot e \times e ,
\end{equation}
so, the non-zero components of the flavour metric are $g_{e \omega} = g_{ \omega e} = -1$. Then, using Eq.\eqref{47} the conserved charge \eqref{41} for Einstein gravity takes the following form
\begin{equation}\label{51}
    Q _{E} ( \xi _{m} ^{\pm} )  =  \frac{1}{8 \pi G} \lim _{r \rightarrow \infty} \int_{0}^{2 \pi} (\xi _{m} ^{\pm})_{a} \left( \delta \omega ^{a} _{\hspace{1.5 mm} \phi} \pm \frac{1}{l} \delta e ^{a} _{\hspace{1.5 mm} \phi} \right) d \phi ,
\end{equation}
In the other hand the connections correspond to the two  $SO(2,1)$ gauge group can be defined as \cite{13}
\begin{equation}\label{52}
  (A^{\pm})^{a} = \omega ^{a} \pm \frac{1}{l} e ^{a} ,
\end{equation}
so, \eqref{51} can be written as
\begin{equation}\label{53}
    Q _{E} ( \xi _{m} ^{\pm} )  =\frac{1}{8 \pi G} \lim _{r \rightarrow \infty} \int_{0}^{2 \pi} (\xi _{m} ^{\pm}) \cdot \delta A^{\pm} _{\hspace{1.5 mm} \phi} d \phi .
\end{equation}
For BTZ black hole spacetime \cite{14}, at spatial infinity we will have
\begin{equation}\label{54}
    \delta e^{a} _{\hspace{1.5 mm} \phi} = 0, \hspace{1 cm} \delta \omega ^{\hat{t}} _{\hspace{1.5 mm} \phi} =-\frac{r_{+}^{2}+r_{-}^{2}}{2 l r} , \hspace{1 cm} \delta \omega ^{\hat{r}} _{\hspace{1.5 mm} \phi} = 0, \hspace {1 cm} \delta \omega ^{\hat{\phi}} _{\hspace{1.5 mm} \phi} =-\frac{r_{+} r_{-}}{ l r},
\end{equation}
By substituting \eqref{45} and \eqref{54} into \eqref{53} we will find the following result
\begin{equation}\label{55}
    Q _{E} ( \xi _{m} ^{\pm} )  = \frac{l}{16 G} \left( \frac{r_{+} \mp r_{-}}{ l } \right) ^{2} \delta _{m,0} .
\end{equation}
In a similar way, for this theory, \eqref{42} will reduced to
\begin{equation}\label{56}
    \delta _{\xi _{n} ^{\pm}} Q _{E} ( \xi _{m} ^{\pm} )  =  \frac{1}{8 \pi G} \lim _{r \rightarrow \infty} \int_{0}^{2 \pi} (\xi _{m} ^{\pm}) \cdot \delta _{\xi _{n} ^{\pm}} A^{\pm} _{\hspace{1.5 mm} \phi} d \phi.
\end{equation}
Then using Eq.\eqref{42} and \eqref{49}, the above equation reduce to the following experssion
\begin{equation}\label{57}
    \delta _{\xi _{n} ^{\pm}} Q _{E} ( \xi _{m} ^{\pm} )  =  \frac{iln^{3}}{8 G} \delta _{m+n,0} .
\end{equation}
Now, by substituting \eqref{55} and \eqref{57} into \eqref{39} we will find the following expression for the central extension term in the considered theory
\begin{equation}\label{58}
  C _{E} (\xi _{m} ^{\pm} , \xi _{n} ^{\pm}) = i \frac{l}{8 G} \left[ n^{3} - \left( \frac{r_{+} \mp r_{-}}{ l } \right) ^{2} n  \right] \delta _{m+n,0} .
\end{equation}
Although this result is in agreement with previous results( for instance, see \cite{15}) but to obtain the usual $n$ dependence, that is $n (n ^{2} -1)$, it is sufficient one make a shift on $Q$ by a constant \cite{16}. Thus, by considering $Q (\xi _{n} ^{\pm}) \equiv L^{\pm} _{n}$ and $ \{ Q ( \xi _{m} ^{\pm} ) , Q ( \xi _{n} ^{\pm} ) \} \equiv i [L^{\pm} _{m} , L^{\pm} _{n}]$, \eqref{36} becomes
\begin{equation}\label{59}
  [L^{\pm} _{m} , L^{\pm} _{n}] =(n-m) L^{\pm} _{m+n} + \frac{c_{\pm}}{12} n (n ^{2} -1) \delta _{m+n,0} ,
\end{equation}
where $c_{\pm}= \frac{3l}{2G}$ are the central charges, and $L^{\pm} _{n}$ are generators of the Virasoro algebra. So the algebra among the conserved charges is isomorphic to two copies of the Virasoro algebra.

\subsection{Generalized massive gravity}
In the Generalized massive gravity (GMG), there are four flavours of one-form, $a^{s}= \{ e, \omega , h, f \}$ and the non-zero components of the flavour metric are \cite{1}
\begin{equation}\label{60}
 g_{e \omega}=-\sigma, \hspace{1 cm} g_{e h}=1, \hspace{1 cm} g_{\omega f}=-\frac{1}{m^{2}}, \hspace{1 cm} g_{\omega \omega}=\frac{1}{\mu}.
\end{equation}
By solving the corresponding equations of motion one can find the following expressions for the auxiliary fields
\begin{equation}\label{61}
  h ^{a} _{\hspace{1.5 mm} \mu} = - \frac{1}{ \mu } S^{a} _{\hspace{1.5 mm} \mu} - \frac{1}{m^{2}} C^{a} _{\hspace{1.5 mm} \mu}, \hspace{1.5 cm} f ^{a}_{\hspace{1.5 mm} \mu} = - S ^{a} _{\hspace{1.5 mm} \mu},
\end{equation}
where $S_{\mu \nu}$ and $C _{\mu \nu}$ are the 3D Schouten tensor and the Cotton tensor, respectively. So, we can obtain for the AdS$_{3}$ background \begin{equation}\label{62}
\begin{split}
     & R^{a}= - \frac{1}{2 l^{2}} (e \times e) ^{a} , \hspace{1 cm} S^{a} = - \frac{1}{2 l^{2}} e^{a} , \hspace{1 cm} C^{a} = 0 , \\
     & \hspace{1 cm} h^{a} =  \frac{1}{2 \mu l^{2}} e^{a} , \hspace{1.5 cm} f^{a} =  \frac{1}{2 l^{2}} e^{a}.
\end{split}
\end{equation}
One can show that \eqref{40} have the following form for this model
\begin{equation}\label{63}
\begin{split}
   Q_{GMG} (\xi _{n} ^{\pm}) = & \left( \sigma \pm \frac{1}{\mu l} +\frac{1}{2 m^{2} l^{2}}  \right) Q_{E} (\xi _{n} ^{\pm})  \\
     & + \frac{1}{8 \pi G} \lim _{r \rightarrow \infty} \int_{0}^{2 \pi} d \phi (\delta h ^{\mu}_{ \hspace{1.5 mm} \phi} \pm \frac{1}{m^{2} l} \delta f ^{\mu}_{ \hspace{1.5 mm} \phi} ) (\xi _{n} ^{\pm})_{\mu}
\end{split}
\end{equation}
 Since the 3D Schouten tensor is given by $S_{\mu \nu} = \mathcal{R} _{\mu \nu} - \frac{1}{4} g _{_{\mu \nu}} \mathcal{R}$ then we have
\begin{equation}\label{64}
  \delta S _{\mu \nu} = \delta \mathcal{R} _{\mu \nu} - \frac{1}{4} g_{\mu \nu} \delta \mathcal{R} + \frac{3}{2 l^{2}} \delta g _{\mu \nu},
\end{equation}
where
\begin{equation}\label{65}
  \begin{split}
       & \delta \mathcal{R} _{\mu \nu} = \frac{1}{2} \left( - \Box \delta g _{\mu \nu} - \nabla _{\mu} \nabla _{\nu} (g^{\alpha \beta} \delta g _{\alpha \beta}) + \nabla ^{\lambda} \nabla _{\mu} \delta g _{\lambda \nu} + \nabla ^{\lambda} \nabla _{\nu} \delta g _{\lambda \mu}  \right), \\
       & \delta \mathcal{R} = - \Box (g^{\alpha \beta} \delta g _{\alpha \beta}) + \nabla ^{\mu} \nabla ^{\nu} \delta g _{\mu \nu} + \frac{2}{l^{2}} (g^{\alpha \beta} \delta g _{\alpha \beta}) ,
  \end{split}
\end{equation}
also, we know that the Cotton tensor is defined as $C^{\mu} _{\hspace{1.5 mm} \nu} =  \epsilon _{\nu} ^{ \hspace{1.5 mm} \alpha \beta} \nabla _{\alpha} S ^{\mu} _{\beta} $ so we can show that its variation is $ \delta C^{\mu} _{\hspace{1.5 mm} \nu} =  \epsilon _{\nu} ^{ \hspace{1.5 mm} \alpha \beta} \nabla _{\alpha} \delta S ^{\mu} _{ \hspace{1.5 mm} \beta} $. For the BTZ black hole solution at the spatial infinity we have
\begin{equation}\label{66}
 \delta g _{t t} = \frac{r_{+}^{2} + r_{-}^{2}}{l^{2}} , \hspace{1 cm} \delta g _{t \phi} = - \frac{r_{+} r_{-}}{l} , \hspace{1 cm} \delta g _{r r} = \frac{ l^{2} (r_{+}^{2} + r_{-}^{2}) }{r^{4}} ,
\end{equation}
then, $\delta C^{\mu} _{\hspace{1.5 mm} \phi} = \delta S ^{\mu} _{\hspace{1.5 mm} \phi} = 0$. Therefore, for the BTZ solution, \eqref{63} will reduce to
\begin{equation}\label{67}
   Q_{GMG} (\xi _{n} ^{\pm}) =  \left( \sigma \pm \frac{1}{\mu l} +\frac{1}{2 m^{2} l^{2}}  \right) Q_{E} (\xi _{n} ^{\pm})
\end{equation}
which show that the conserved charges of BTZ solution of GMG are simply a constant time the conserved charge of BTZ solution of Einstein gravity.
In a similar way, for GMG model, we can show that \eqref{42} can be simplify as
\begin{equation}\label{68}
\begin{split}
   \delta _{\xi _{m} ^{\pm}} Q_{GMG} (\xi _{n} ^{\pm}) = & \left( \sigma \pm \frac{1}{\mu l} +\frac{1}{2 m^{2} l^{2}}  \right) \delta _{\xi _{m} ^{\pm}} Q_{E} (\xi _{n} ^{\pm})  \\
     & + \frac{1}{8 \pi G} \lim _{r \rightarrow \infty} \int_{0}^{2 \pi} d \phi (\delta _{\xi _{m} ^{\pm}} h ^{\mu}_{ \hspace{1.5 mm} \phi} \pm \frac{1}{m^{2} l} \delta _{\xi _{m} ^{\pm}} f ^{\mu}_{ \hspace{1.5 mm} \phi} ) (\xi _{n} ^{\pm})_{\mu}.
\end{split}
\end{equation}
One can show that $\delta _{\xi _{m} ^{\pm}} h^{\mu} _{\hspace{1.5 mm} \phi} = \delta _{\xi _{m} ^{\pm}} f ^{\mu} _{\hspace{1.5 mm} \phi} = 0$, so \eqref{68} can be rewritten as
\begin{equation}\label{69}
   \delta _{\xi _{m} ^{\pm}} Q_{GMG} (\xi _{n} ^{\pm}) =  \left( \sigma \pm \frac{1}{\mu l} +\frac{1}{2 m^{2} l^{2}}  \right) \delta _{\xi _{m} ^{\pm}} Q_{E} (\xi _{n} ^{\pm}).
\end{equation}
Now, by substituting \eqref{67} and \eqref{69} into \eqref{39}, we find that
\begin{equation}\label{70}
  C _{GMG} (\xi _{m} ^{\pm} , \xi _{n} ^{\pm}) = \left( \sigma \pm \frac{1}{\mu l} +\frac{1}{2 m^{2} l^{2}}  \right) C _{E} (\xi _{m} ^{\pm} , \xi _{n} ^{\pm}) ,
\end{equation}
so, we can easily read off the central charges of the General Massive Gravity as follows:
\begin{equation}\label{71}
  c_{\pm} = \frac{3l}{2G} \left( \sigma \pm \frac{1}{\mu l} +\frac{1}{2 m^{2} l^{2}}  \right).
\end{equation}
This result is in agreement with what was found in \cite{17}. We can read of the eigenvalues of the Virasoro generators $L^{\pm} _{n}$ from \eqref{67} as
\begin{equation}\label{72}
   l^{\pm} _{n} = \frac{l}{16 G} \left( \sigma \pm \frac{1}{\mu l} +\frac{1}{2 m^{2} l^{2}}  \right) \left( \frac{r_{+} \mp r_{-}}{ l } \right) ^{2} \delta _{n,0},
\end{equation}
The eigenvalues of the Virasoro generators $L ^{\pm} _{n} $ are related to the energy $E$ and the angular momentum $j$ of the BTZ black hole by the following equations respectively
\begin{equation}\label{73}
   E = l^{-1} ( l^{+} _{0} + l^{-} _{0} ) = \frac{1}{8 G} \left[ \left( \sigma +\frac{1}{2 m^{2} l^{2}}  \right) \frac{r_{+}^{2}+r_{-}^{2} }{l^{2}} - \frac{2 r_{+} r_{-}}{\mu l^{3}} \right] ,
\end{equation}
\begin{equation}\label{74}
   j = l^{-1} (l^{+} _{0} - l^{-} _{0}) =  \frac{1}{8 G} \left[ \left( \sigma +\frac{1}{2 m^{2} l^{2}}  \right) \frac{2 r_{+} r_{-}}{ l }  - \frac{r_{+}^{2}+r_{-}^{2} }{ \mu l^{2}} \right] .
\end{equation}
Also, we can calculate the entropy of the considered black hole solution by the Cardy formula  \cite{18,19} (see also \cite{12} )
\begin{equation}\label{75}
   S = 2 \pi \sqrt{\frac{c_{+} l^{+} _{0} }{6}} + 2 \pi \sqrt{\frac{c_{-} l^{-} _{0} }{6}} ,
\end{equation}
then using Eqs.(\ref{71}),(\ref{72}), we will have
\begin{equation}\label{76}
   S = \frac{\pi}{2 G} \left[ \left( \sigma +\frac{1}{2 m^{2} l^{2}}  \right) r_{+} - \frac{r_{-}}{\mu l} \right] .
\end{equation}
Now, we try to find out the energy, the angular momentum and the entropy of the BTZ black hole from conserved charge formula \eqref{34} by the corresponding vector fields. To this end, we suppose that $s=0$ and $s=1$ are correspond to the AdS$_{3}$ background and the BTZ black hole spacetime, respectively. Thus, for the General Massive Gravity \eqref{33} takes following form
\begin{equation}\label{77}
    \delta Q ( \xi ) = \frac{1}{8 \pi G} \int_{0}^{2 \pi} d \phi \left[ \left( \sigma +\frac{1}{2 m^{2} l^{2}}  \right) \xi _{a} + \frac{1}{\mu } \Xi _{a} \right] \cdot \delta \omega ^{a} _{\hspace{1.5 mm} \phi} ,
\end{equation}
where the integration runs over a circle of arbitrary radii also. In the above equation $\xi^{a}$ and $ \Xi ^{a} $ are given by
\begin{equation}\label{78}
   \xi^{a} = e ^{a} _{\hspace{1.5 mm} \mu} \xi ^{\mu} , \hspace{1.5 cm} \Xi ^{a} = - \frac{1}{2} e ^{a} _{\hspace{1.5 mm} \lambda } \epsilon ^{\lambda \mu \nu} \nabla _{\mu} \xi _{\nu} .
\end{equation}
The energy, angular momentum and the entropy of the BTZ black hole are correspond to the following Killing vectors respectively
\begin{equation}\label{79}
   \xi _{(E)} = \partial _{t} , \hspace{1 cm} \xi _{(j)} = - \partial _{\phi} , \hspace{1 cm} \xi_ {(S)} = \frac{4 \pi}{\kappa} ( \partial _{t} + \Omega _{H} \partial _{\phi} ),
\end{equation}
where $\Omega _{H}=\frac{r_{-}}{l r_{+}}$ is the angular velocity of horizon and $\kappa = \frac{r_{+}^{2}-r_{-}^{2}}{l^{2} r_{+}}$ is the surface gravity. By substituting these killing vectors into \eqref{77} and making an integration over an one-parameter path on the solution space, we will find \eqref{73}, \eqref{74} and \eqref{76} exactly. It is easy to show that these results satisfy the first law of black hole mechanics.

\section{Conclusion}
In this paper we have considered the Chern-Simons-like theories of gravity (CSLTG) in the context of the first order formalism. We have studied the problem of defining off-shell conserved charges in the framework of the CSLTG. In order to obtain the ADT current, we used the formalism presented in \cite{2}. We know that the ADT current is an off-shell current and has defined by the virtue of the Killing vector fields. We have generalized the ADT current such that it is conserve for any diffeomorphism generator vector fields. We have fixed the ambiguity in definition of the generalized ADT conserved current by considering the phase space analysis. Form this generalized ADT conserved current, in section 4, we read off the conserved charge associated with a diffeomorphism generator vector field $\xi$ by Eq.\eqref{34} and its perturbation by Eq.\eqref{33}. We have shown that the perturbation of the conserved charge is off-shell and is conserved for any diffeomorphism generator $\xi$. On the other hand, since these conserved charges are quasi-local so we can consider the asymptotic symmetries as well as near horizon symmetries to obtain central extension term for any black hole solution. In section 5, we have considered a set of vector fields which are satisfy the Witt algebra and have found a general expression for the central extension term at spatial infinity (see Eqs.\eqref{39}, \eqref{41} and \eqref{42}). Since the quasi-local conserved charge perturbation \eqref{33} is independent of the integration surface and is conserved for any diffeomorphism generator $\xi$, then the one we found by Eq.\eqref{34} holds near horizon. In section 6, we have applied the method to the Einstein gravity with negative cosmological constant and the Generalized Massive Gravity as examples. For these examples, we have calculated the central extension term and, through it, we read off the central charges and the eigenvalues of the Virasoro algebra generators for the BTZ black hole solution. We have obtained the entropy of the BTZ black hole \eqref{76} by using the Cardy formula.  Also we have calculated the energy \eqref{73} and the angular momentum \eqref{74} of this black hole using the eigenvalues of the Virasoro algebra generators. Eventually, we have shown that the formula \eqref{34} gives the same results.

\section{Acknowledgments}
M. R. Setare  thanks Dr. A. Sorouri  for his help in improvement the English of the text.

\end{document}